\documentclass{fundam}

\usepackage{amssymb}
\usepackage{euscript}
\usepackage{graphicx}

\newcommand{\ac}[1]{c:\Sigma\times\Sigma^{\leq{#1}}\rightarrow\Delta^{+}}
\newcommand{\eac}{\overline{c}:\Sigma^{*}\rightarrow\Delta^{*}}
\newcommand{\sstring}[1]{\sigma_{1}\sigma_{2}\ldots\sigma_{#1}}
\newcommand{\ttup}[1]{\texttt{\textup{#1}}}

\begin{document}
%\setcounter{page}{1001}
%\issue{XXI~(2001)}

\title{Special Cases of Encodings by Generalized Adaptive Codes}

\author{Drago\c s Trinc\u a\\
Department of Computer Science \& Engineering\\
University of Connecticut\\
371 Fairfield Road, Unit 2155, Storrs, CT 06269-2155, USA\\
dnt04001{@}engr.uconn.edu}

\maketitle

\runninghead{Drago\c s Trinc\u a}{Special Cases of Encodings by Generalized Adaptive Codes}

\begin{abstract}
	Adaptive (variable-length) codes associate variable-length codewords
	to symbols being encoded depending on the previous symbols in the input data string.
	This class of codes has been presented in \cite{t:1,t:2} as a new class of non-standard variable-length codes.
	Generalized adaptive codes (GA codes, for short) have been also presented in \cite{t:1,t:2} not only as a
	new class of non-standard variable-length codes, but also as a natural generalization of adaptive codes
	of any order. This paper is intended to continue developing the theory of variable-length codes by establishing several interesting connections
	between adaptive codes and other classes of codes. The connections are discussed not only from a theoretical point of view
	(by proving new results), but also
	from an applicative one (by proposing several applications). First, we prove that adaptive Huffman encodings and Lempel-Ziv encodings
	are particular cases of encodings by GA codes. Second, we show that any $(n,1,m)$ convolutional code satisfying certain conditions can
	be modelled as an adaptive code
	of order $m$. Third, we describe a cryptographic scheme based on the connection between adaptive codes and convolutional codes,
	and present an insightful analysis of this scheme.
	Finally, we conclude by generalizing adaptive codes to $(p,q)$-adaptive codes, and discussing connections
	between adaptive codes and time-varying codes.
\end{abstract}


\section{Introduction}
	The theory of variable-length codes, one of the most studied areas of coding theory, continues to play an
	important role not only in the evolution of formal languages, but also in some applicative areas of computer science such as
	data compression. The aim of this paper is to continue developing and enriching this theory with new results, along with
	showing their effectiveness in concrete applications.
	
	Specifically, we continue our study on {\it adaptive codes}, which have been recently presented in \cite{t:1,t:2} as a new class
	of non-standard
	variable-length codes. Intuitively, an adaptive code of order $n$ associates a codeword to
	the symbol being encoded depending on the previous $n$ symbols in the input data string.
	Generalized adaptive codes (GA codes, for short) have been also presented in \cite{t:1,t:2} not only as a
	new class of non-standard variable-length codes, but also as a natural generalization of adaptive codes
	of any order.

	Both classes are described in detail in section 2.
	Then, we show that adaptive Huffman encodings and Lempel-Ziv encodings
	are particular cases of encodings by GA codes (sections 3 and 4).
	In section 5, we show that any $(n,1,m)$ convolutional code satisfying a certain condition can be modelled as an adaptive code
	of order $m$.
	This result is exploited further in section 6, where an efficient cryptographic scheme based on convolutional codes is described.
	An insightful analysis of this cryptographic scheme is provided in the same section.
	In sections 7 and 8, we extend adaptive codes to $(p,q)$-adaptive codes, and present a new class of variable-length codes, called
	adaptive time-varying codes.
	
	In the remainder of this introductory section, we recall some basic notions and notations used
	throughout the paper.
	We denote by $|S|$ the \textit{cardinality} of the set $S$; if $x$ is a string of
	finite length, then $|x|$ denotes the length of $x$.
	The \textit{empty string} is denoted by $\lambda$.

	For an alphabet $\Sigma$,
	we denote by $\Sigma^{*}$ the set
	$\bigcup_{n=0}^{\infty}\Sigma^{n}$
	and by $\Sigma^{+}$ the set
	$\bigcup_{n=1}^{\infty}\Sigma^{n}$,
	where $\Sigma^{0}$ is the set $\{\lambda\}$.
	Also, we denote by
	$\Sigma^{\leq n}$ the set
	$\bigcup_{i=0}^{n}\Sigma^{i}$
	and by $\Sigma^{\geq n}$ the set
	$\bigcup_{i=n}^{\infty}\Sigma^{i}$.
	Let us consider an alphabet $\Delta$, $X$ a finite and nonempty subset of $\Delta^{+}$, and $w\in\Delta^{+}$.
	A \textit{decomposition of} $w$ over $X$ is any sequence of strings
	$u_{1}, u_{2}, \ldots, u_{h}$ with $u_{i}\in X$ for all $i$, $1\leq i\leq h$, 
	such that $w=u_{1}u_{2}\ldots u_{h}$.
	A \textit{code} over $\Delta$ is any nonempty set $C\subseteq\Delta^{+}$ such
	that each string $w\in\Delta^{+}$ has at most one decomposition over $C$.
	A \textit{prefix code} over $\Delta$ is any code $C$ over $\Delta$ such that
	no string in $C$ is proper prefix of another string in $C$. 
	
	If $\mathcal{A}$ is an algorithm and $x$ its input, then we denote by $\mathcal{A}(x)$ its output.
	Also, we denote by $\mathbb{N}$ the set of natural numbers, and by $\mathbb{N}^{*}$ the set of nonzero natural numbers.
	
	Finally, let us fix some useful notations which will be used in the description of the algorithms.
	Let $\mathcal{U}=(u_{1},u_{2},\ldots,u_{k})$ be a $k$-tuple.
	We denote by $\mathcal{U}.i$ the $i$-th component of $\mathcal{U}$,
	that is, $\mathcal{U}.i=u_{i}$ for all $i\in\{1,2,\ldots,k\}$.
	The $0$-tuple is denoted by $()$. The length of a tuple $\mathcal{U}$ is denoted by ${\it Len}(\mathcal{U})$.
	If $\mathcal{V}=(v_{1},v_{2},\ldots,v_{b})$,
	$\mathcal{M}=(m_{1},m_{2},\ldots,m_{r},\mathcal{U})$, $\mathcal{N}=(n_{1},n_{2},\ldots,n_{s},\mathcal{V})$,
	and $\mathcal{P}=(p_{1},\ldots,p_{i-1},p_{i},p_{i+1},\ldots,p_{t})$ are tuples, and q is an element or a tuple,
	then we define
	$\mathcal{P}\vartriangleleft{q}$, $\mathcal{P}\vartriangleright{i}$, $\mathcal{U}\vartriangle{\mathcal{V}}$, and
	$\mathcal{M}\lozenge{\mathcal{N}}$ by:
\begin{itemize}
\item $\mathcal{P}\vartriangleleft{q}=(p_{1},\ldots,p_{t},q)$,
\item $\mathcal{P}\vartriangleright{i}=(p_{1},\ldots,p_{i-1},p_{i+1},\ldots,p_{t})$,
\item $\mathcal{U}\vartriangle{\mathcal{V}}=(u_{1},u_{2},\ldots,u_{k},v_{1},v_{2},\ldots,v_{b})$,
\item $\mathcal{M}\lozenge{\mathcal{N}}=(m_{1}+n_{1},m_{2}+1,\ldots,m_{r}+1,n_{2}+1,\ldots,n_{s}+1,\mathcal{U}\vartriangle{\mathcal{V}})$,
\end{itemize}
	where $m_{1},m_{2},\ldots,m_{r},n_{1},n_{2},\ldots,n_{s}$ are integers.

\section{Adaptive codes and GA codes}
	The aim of this section is to briefly review some basic
	definitions, results, and notations related to adaptive codes and generalized adaptive codes \cite{t:1,t:2}.
\begin{definition}
\label{def:ac}
	Let $\Sigma$ and $\Delta$ be two alphabets. A function
	$\ac{n}$, $n\geq{1}$, is called \emph{adaptive code of order $n$} if its unique
	homomorphic extension $\eac$, given by:
\begin{itemize}
\item $\overline{c}(\lambda)=\lambda$,
\item $\overline{c}(\sstring{m})=$
	$c(\sigma_{1},\lambda)$
	$c(\sigma_{2},\sigma_{1})$
	$\ldots$
	$c(\sigma_{n-1},\sstring{n-2})$
	\newline
	$c(\sigma_{n},\sstring{n-1})$
	$c(\sigma_{n+1},\sstring{n})$
	$c(\sigma_{n+2},\sigma_{2}\sigma_{3}\ldots\sigma_{n+1})$
	\newline
	$c(\sigma_{n+3},\sigma_{3}\sigma_{4}\ldots\sigma_{n+2})\ldots$
	$c(\sigma_{m},\sigma_{m-n}\sigma_{m-n+1}\ldots\sigma_{m-1})$
\end{itemize}
	for all strings $\sstring{m}\in\Sigma^{+}$, is injective.
\end{definition}
	As it is clearly specified in the definition above, an adaptive code of order $n$ associates a variable-length
	codeword to the symbol being encoded depending on the previous $n$ symbols in the input data string.
	Let us take an example in order to better understand this mechanism.
\begin{example}
	Let $\Sigma=\{\ttup{a},\ttup{b}\}$ and $\Delta=\{0,1\}$ be two alphabets, and
	$\ac{1}$ a function given as in the table below.
	One can verify that $\overline{c}$ is injective, and according to Definition \ref{def:ac}, it follows that $c$ is
	an adaptive code of order one.
\begin{table}[htbp]
\caption{An adaptive code of order one.}
\begin{center}
\begin{tabular}{|c|c|c|c|}
\hline
$\Sigma\backslash\Sigma^{\leq{1}}$	& $\ttup{a}$ & $\ttup{b}$	& $\lambda$	\\ \hline
$\ttup{a}$ 					        & 0          & 1 	        & 00	   	\\ \hline
$\ttup{b}$					        & 10         & 00	        & 11       	\\ \hline
\end{tabular}	
\end{center}
\end{table}
\hspace{0pt}
\newline
	Let $x=\ttup{abaa}\in\Sigma^{+}$ be an input data string. 
	Using the definition above, we encode $x$ by
\begin{center}
	$\overline{c}(x)=c(\ttup{a},\lambda)c(\ttup{b},\ttup{a})c(\ttup{a},\ttup{b})c(\ttup{a},\ttup{a})=001010$.
\end{center}
\end{example}
\begin{example}
	Let us consider $\Sigma=\{\ttup{a},\ttup{b},\ttup{c}\}$ and $\Delta=\{0,1\}$ two alphabets, and
	$\ac{2}$ a function given as in the following table.
	One can easily verify that $\overline{c}$ is injective, and according to Definition \ref{def:ac}, $c$ is
	an adaptive code of order two.
\begin{table}[htbp]
\caption{An adaptive code of order two.}
\begin{center}
\begin{tabular}{|c|c|c|c|c|c|c|c|c|c|c|c|c|c|}
\hline
$\Sigma\backslash\Sigma^{\leq{2}}$	& $\ttup{a}$ & $\ttup{b}$	& $\ttup{c}$	& $\ttup{aa}$ & $\ttup{ab}$ & $\ttup{ac}$ & $\ttup{ba}$ & $\ttup{bb}$ & $\ttup{bc}$ & $\ttup{ca}$ & $\ttup{cb}$ & $\ttup{cc}$ & $\lambda$	\\ \hline
$\ttup{a}$ 					        & 0          & 11 	        & 10 	        & 00 	      & 1           & 10          & 01          & 10          & 11          & 11          & 11          & 0           & 00	   		\\ \hline
$\ttup{b}$					        & 10         & 000	        & 11	        & 11	      & 01          & 00          & 00          & 11          & 01          & 101         & 00          & 10          & 11       	\\ \hline
$\ttup{c}$					        & 111        & 01	        & 00	        & 10	      & 00          & 11          & 11          & 00          & 00          & 0           & 10          & 11          & 10	   		\\ \hline
\end{tabular}	
\end{center}
\end{table}
\hspace{0pt}
\newline
	Let $x=\ttup{abacca}\in\Sigma^{+}$ be an input data string. 
	Using the definition above, we encode $x$ by
\begin{center}
	$\overline{c}(x)=c(\ttup{a},\lambda)c(\ttup{b},\ttup{a})c(\ttup{a},\ttup{ab})c(\ttup{c},\ttup{ba})c(\ttup{c},\ttup{ac})c(\ttup{a},\ttup{cc})=0010111110$.
\end{center}
\end{example}
	Let $\ac{n}$ be an adaptive code of order $n$, $n\geq{1}$. We denote by
	$C_{c, \sigma_{1}\sigma_{2}\ldots\sigma_{h}}$ the set
	$\{c(\sigma,\sigma_{1}\sigma_{2}\ldots\sigma_{h}) \mid \sigma\in\Sigma\}$,
	for all $\sigma_{1}\sigma_{2}\ldots\sigma_{h}\in\Sigma^{\leq{n}}-\{\lambda\}$,
	and by $C_{c, \lambda}$ the set $\{c(\sigma,\lambda) \mid \sigma\in\Sigma\}$.
	We write $C_{\sigma_{1}\sigma_{2}\ldots\sigma_{h}}$ instead of
	$C_{c, \sigma_{1}\sigma_{2}\ldots\sigma_{h}}$,
	and $C_{\lambda}$ instead of $C_{c, \lambda}$
	whenever there is no confusion.
	Let us denote by $AC(\Sigma,\Delta,n)$ the set
\begin{center}
	$\{\ac{n} \mid c$ is an adaptive code of order $n\}$.
\end{center}
\begin{theorem}
\label{thm:1}
	Let $\Sigma$ and $\Delta$ be two alphabets, and $\ac{n}$ a function, $n\geq{1}$.
	If $C_{u}$ is prefix code, for all $u\in\Sigma^{\leq{n}}$, then
	$c\in{AC(\Sigma,\Delta,n)}$.
\end{theorem}
\begin{proof}
	Let us assume that $C_{\sstring{h}}$ is prefix code, for all
	$\sstring{h}\in\Sigma^{\leq{n}}$, but $c\notin{AC(\Sigma,\Delta,n)}$.
	By Definition \ref{def:ac}, the unique homomorphic extension of $c$, denoted by $\overline{c}$,
	is not injective.
	This implies that $\exists$ $u\sigma u', u\sigma'u''\in\Sigma^{+}$, with
	$\sigma,\sigma '\in\Sigma$ and $u,u',u''\in\Sigma^{*}$, such that
	$\sigma\neq\sigma'$ and
\begin{equation}
	\overline{c}(u\sigma u')=\overline{c}(u\sigma'u'').
\end{equation}
	We can rewrite the equality (1) by
\begin{equation}
	\overline{c}(u)c(\sigma,{P}_{n}(u))\overline{c}(u')=
	\overline{c}(u)c(\sigma',{P}_{n}(u))\overline{c}(u''),
\end{equation}
	where the function ${P}_{n}(\cdot)$ is given as below.
\begin{displaymath}
{P}_{n}(u)=
\left\{ 	
	\begin{array}{ll}
	\lambda 			& \textrm{if $u=\lambda$.} \\
	u_{1}\ldots u_{q} 	& \textrm{if $u=u_{1}u_{2}\ldots u_{q}$ and $u_{1},u_{2},\ldots,u_{q}\in\Sigma$
							and $q\leq{n}$.} \\
	u_{q-n+1}\ldots u_{q} 	& \textrm{if $u=u_{1}u_{2}\ldots u_{q}$ and $u_{1},u_{2},\ldots,u_{q}\in\Sigma$
							and $q>n$.}
	\end{array} 
\right.
\end{displaymath}
	By hypothesis, $C_{{P}_{n}(u)}$ is prefix code and
	$c(\sigma,{P}_{n}(u)),c(\sigma',{P}_{n}(u))\in{C_{{P}_{n}(u)}}$.

	Therefore, the set $\{c(\sigma,{P}_{n}(u)),c(\sigma',{P}_{n}(u))\}$ is a prefix code.
	But the equality (2) holds true if and only if
	$\{c(\sigma,{P}_{n}(u)),c(\sigma',{P}_{n}(u))\}$ is not a prefix set.
	Thus, our assumption leads to a contradiction.
\end{proof}
\begin{definition}
\label{def:gac}
	Let $F:\mathbb{N}^{*}\times\Sigma^{+}\rightarrow\Sigma^{*}$ be a function, where $\mathbb{N}^{*}$ denotes the set $\mathbb{N}-\{0\}$.
	A function
	$c_{F}:\Sigma\times\Sigma^{*}\rightarrow\Delta^{+}$ is called
	\emph{generalized adaptive code} (GA code, for short) if its unique homomorphic extension
	$\overline{c_{F}}:\Sigma^{*}\rightarrow\Delta^{*}$, given by:
\begin{itemize}
\item $\overline{c_{F}}(\lambda)=\lambda$,
\item $\overline{c_{F}}(\sstring{m})=c_{F}(\sigma_{1},F(1,\sstring{m}))\ldots{c_{F}(\sigma_{m},F(m,\sstring{m}))}$
\end{itemize}
	for all strings $\sstring{m}\in\Sigma^{+}$, is injective.
\end{definition}
\begin{remark}
\label{rmk:1}
	The function $F$ in Definition \ref{def:gac} is called the \emph{adaptive function}
	corresponding to the GA code $c_{F}$.
	Clearly, a GA code $c_{F}$ can be constructed if its adaptive function $F$
	is already constructed.
\end{remark}
\begin{remark}
\label{rmk:2}
	Let $\Sigma$ and $\Delta$ be two alphabets. We denote by $GAC(\Sigma,\Delta)$ the set
\begin{center}
	$\{c_{F}:\Sigma\times\Sigma^{*}\rightarrow\Delta^{+}$ $\mid$ $c_{F}$ is a GA code$\}$.
\end{center}
\end{remark}
	The following theorem proves that adaptive codes (of any order) are special cases
	of GA codes.
\begin{theorem}
\label{thm:2}
	Let $\Sigma$ and $\Delta$ be alphabets. Then,
	$AC(\Sigma,\Delta,n)\subseteq{GAC(\Sigma,\Delta)}$
	for all $n\geq{1}$.
\end{theorem}
\begin{proof}
	Let $c_{F}\in{AC(\Sigma,\Delta,n)}$ be an adaptive code of order $n$, $n\geq{1}$,
	and $F:\mathbb{N}^{*}\times\Sigma^{+}\rightarrow\Sigma^{*}$ a function given by:
\begin{displaymath}
F(i,\sstring{m})=
\left\{ 	
	\begin{array}{ll}
	\lambda 			
					& \textrm{if $i=1$ or $i>m$.}		 			\\
	\sstring{i-1}		
					& \textrm{if $2\leq{i}\leq{m}$ and $i\leq{n+1}$.} 	\\
	\sigma_{i-n}\sigma_{i-n+1}\ldots\sigma_{i-1} 	
					& \textrm{if $2\leq{i}\leq{m}$ and $i>n+1$.}		
	\end{array} 
\right.
\end{displaymath}
	for all $i\geq{1}$ and $\sstring{m}\in{\Sigma^{+}}$.
	One can verify that $|F(i,\sstring{m})|\leq{n}$, for all 
	$i\geq{1}$ and $\sstring{m}\in{\Sigma^{+}}$.
	According to Definition \ref{def:ac}, the function $\overline{c_{F}}$ is given by:
\begin{itemize}
\item $\overline{c_{F}}(\lambda)=\lambda$,
\item $\overline{c_{F}}(\sstring{m})=$
	$c_{F}(\sigma_{1},\lambda)$
	$c_{F}(\sigma_{2},\sigma_{1})$
	$\ldots$
	$c_{F}(\sigma_{n-1},\sstring{n-2})$
	\newline
	$c_{F}(\sigma_{n},\sstring{n-1})$
	$c_{F}(\sigma_{n+1},\sstring{n})$
	$c_{F}(\sigma_{n+2},\sigma_{2}\sigma_{3}\ldots\sigma_{n+1})$
	\newline
	$c_{F}(\sigma_{n+3},\sigma_{3}\sigma_{4}\ldots\sigma_{n+2})\ldots$
	$c_{F}(\sigma_{m},\sigma_{m-n}\sigma_{m-n+1}\ldots\sigma_{m-1})$
\end{itemize}
	for all strings $\sstring{m}\in\Sigma^{+}$.
	It is easy to remark that
\begin{center}
	$\overline{c_{F}}(\sstring{m})=c_{F}(\sigma_{1},F(1,\sstring{m}))\ldots c_{F}(\sigma_{m},F(m,\sstring{m}))$
\end{center}
	for all $\sstring{m}\in\Sigma^{+}$, which proves the theorem.
\end{proof}
	The adaptive mechanism in Definition \ref{def:gac} can be illustrated by the figure below.
	More precisely, the figure captures the idea behind this mechanism: the codeword associated to the current symbol
	depends on the symbol itself and a sequence of symbols chosen by the adaptive function.
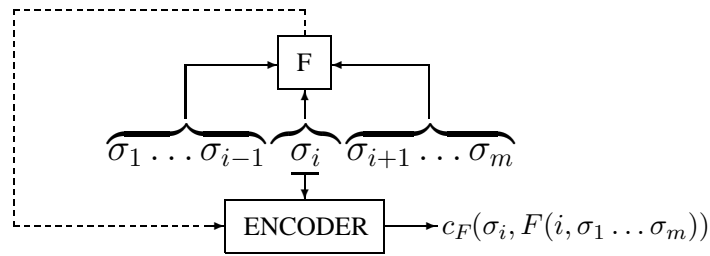
\begin{figure}[hbtp]
\setlength{\unitlength}{1pt}
\begin{picture}(340,100)
			\begin{Large}
			\linethickness{0.4pt}
			\put(130.5,38)
{$\overbrace{\sigma_{1}\ldots\sigma_{i-1}}\overbrace{\sigma_{i}}\overbrace{\sigma_{i+1}\ldots\sigma_{m}}$}
		
			\put(200,33){\line(1,0){10}}
			\put(205,33){\vector(0,-1){10}}
			
			\put(175,23){\line(1,0){60}}
			\put(175,3){\line(1,0){60}}
			\put(175,23){\line(0,-1){20}}
			\put(235,23){\line(0,-1){20}}
			\put(182,10){\small{ENCODER}}
			\put(235,13){\vector(1,0){20}}
			\put(257,10){\normalsize{$c_{F}$}}
			\put(268,10){\normalsize{$(\sigma_{i},F(i, \sigma_{1}\ldots\sigma_{m}))$}}
			
			\put(195,65){\line(1,0){20}}
			\put(195,85){\line(1,0){20}}
			\put(195,65){\line(0,1){20}}
			\put(215,65){\line(0,1){20}}
			\put(202,72){\small{F}}

			\put(205,54){\vector(0,1){11}}
			\put(160,54){\line(0,1){20}}
			\put(252,54){\line(0,1){20}}
			\put(160,74){\vector(1,0){35}}
			\put(252,74){\vector(-1,0){37}}

			\put(205,85){\line(0,1){2}}
			\put(205,89){\line(0,1){2}}
			\put(205,93){\line(0,1){2}}
			
			\put(205,95){\line(-1,0){2}}
			\put(201,95){\line(-1,0){2}}
			\put(197,95){\line(-1,0){2}}
			\put(193,95){\line(-1,0){2}}
			\put(189,95){\line(-1,0){2}}
			\put(185,95){\line(-1,0){2}}
			\put(181,95){\line(-1,0){2}}
			\put(177,95){\line(-1,0){2}}
			\put(173,95){\line(-1,0){2}}
			\put(169,95){\line(-1,0){2}}
			\put(165,95){\line(-1,0){2}}
			\put(161,95){\line(-1,0){2}}\put(157,95){\line(-1,0){2}}\put(153,95){\line(-1,0){2}}
			\put(149,95){\line(-1,0){2}}\put(145,95){\line(-1,0){2}}\put(141,95){\line(-1,0){2}}
			\put(137,95){\line(-1,0){2}}\put(133,95){\line(-1,0){2}}\put(129,95){\line(-1,0){2}}
			\put(125,95){\line(-1,0){2}}\put(121,95){\line(-1,0){2}}\put(117,95){\line(-1,0){2}}
			\put(113,95){\line(-1,0){2}}\put(109,95){\line(-1,0){2}}\put(105,95){\line(-1,0){2}}
			\put(101,95){\line(-1,0){2}}\put(97,95){\line(-1,0){2}}
			\put(95,95){\line(0,-1){2}}
			\put(95,91){\line(0,-1){2}}
			\put(95,87){\line(0,-1){2}}
			\put(95,83){\line(0,-1){2}}
			\put(95,79){\line(0,-1){2}}
			\put(95,75){\line(0,-1){2}}
			\put(95,71){\line(0,-1){2}}
			\put(95,67){\line(0,-1){2}}
			\put(95,63){\line(0,-1){2}}
			\put(95,59){\line(0,-1){2}}
			\put(95,55){\line(0,-1){2}}
			\put(95,51){\line(0,-1){2}}
			\put(95,47){\line(0,-1){2}}
			\put(95,43){\line(0,-1){2}}
			\put(95,39){\line(0,-1){2}}
			\put(95,35){\line(0,-1){2}}
			\put(95,31){\line(0,-1){2}}
			\put(95,27){\line(0,-1){2}}
			\put(95,23){\line(0,-1){2}}
			\put(95,19){\line(0,-1){2}}
			\put(95,15){\line(0,-1){2}}

			\put(95,13){\line(1,0){2}}
			\put(99,13){\line(1,0){2}}
			\put(103,13){\line(1,0){2}}
			\put(107,13){\line(1,0){2}}
			\put(111,13){\line(1,0){2}}
			\put(115,13){\line(1,0){2}}
			\put(119,13){\line(1,0){2}}
			\put(123,13){\line(1,0){2}}
			\put(127,13){\line(1,0){2}}
			\put(131,13){\line(1,0){2}}
			\put(135,13){\line(1,0){2}}
			\put(139,13){\line(1,0){2}}
			\put(143,13){\line(1,0){2}}
			\put(147,13){\line(1,0){2}}
			\put(151,13){\line(1,0){2}}
			\put(155,13){\line(1,0){2}}
			\put(159,13){\line(1,0){2}}
			\put(163,13){\line(1,0){2}}
			\put(167,13){\vector(1,0){8}}
			\end{Large}
\end{picture}
\caption{Encoding with a GA code.}
\end{figure}
\begin{example}
\label{exmp:1}
	Let $\Sigma$ and $\Delta$ be two alphabets, 
	$c_{F}:\Sigma\times\Sigma^{*}\rightarrow\Delta^{+}$ a GA code, and
	$F:\mathbb{N}^{*}\times\Sigma^{+}\rightarrow\Sigma^{*}$ its adaptive function.
	Let us consider $F$ given as below. 
\begin{displaymath}
	F(i,\sstring{m})=
	\left\{
		\begin{array}{ll}
		\lambda		& \textrm{if $i=1$ or $i>m$.}	\\
		\sigma_{i-1}	& \textrm{if $2\leq{i}\leq{m}$.}
		\end{array}
	\right.
\end{displaymath}
	One can trivially verify that the function $c_{F}$ is also an adaptive code of order one.
\end{example}

\section{GA codes and adaptive Huffman codes}
	In this section, we prove that adaptive Huffman encodings are particular cases of encodings by GA codes.
	This result can be exploited further in data compression to develop efficient compression algorithms; for example,
	the algorithms presented in \cite{t:2} combine adaptive codes with Huffman's classical algorithm.
	
	The well-known Huffman algorithm is a two-pass encoding scheme, that is,
	the input must be read twice. The version used in practice is called the \emph{adaptive
	Huffman algorithm}, which reads the input only once. Intuitively, the encoding of an
	input data string using the adaptive Huffman algorithm requires the construction of
	a sequence of {\it Huffman trees}.
\newline\indent
	Let $\Sigma$ be an alphabet, and $w=w_{1}w_{2}\ldots{w_{h}}$ a string over $\Sigma$.
	Denote by $\mathcal{T}_{0}(w),\mathcal{T}_{1}(w),\ldots,\mathcal{T}_{h}(w)$
	the sequence of Huffman trees constructed by the adaptive Huffman algorithm for the input string $w$.
	The Huffman tree $\mathcal{T}_{0}(w)$ is associated to the alphabet $\Sigma$ (with the assumption that
	each symbol in $\Sigma$ has frequency $1$).
	For all $i\in\{1,2,\ldots,h\}$, the Huffman tree $\mathcal{T}_{i}(w)$ (associated to the
	string $w_{1}w_{2}\ldots{w_{i}}$) is obtained by updating the tree $\mathcal{T}_{i-1}(w)$.
\newline\indent	
	The procedure via this update takes place is called the \textit{sibling transformation}, which
	can be described as follows. Let $\mathcal{T}_{i}(w)$ be the current tree and $k$ the frequency of $w_{i+1}$;
	the tree $\mathcal{T}_{i+1}(w)$ is obtained from $\mathcal{T}_{i}(w)$ by applying the following algorithm:
	compare $w_{i+1}$ with its successors in the tree (from left to right and from bottom to top).
	If the immediate successor has frequency $k+1$ or greater, then we do not have to change
	anything. Otherwise, $w_{i+1}$ should be swapped with the last successor which has frequency $k$
	or smaller (only if this successor is not its parent). The frequency of $w_{i+1}$ is incremented
	from $k$ to $k+1$. If $w_{i+1}$ is the root of the tree, then the loop terminates. Otherwise, it
	continues with the parent of $w_{i+1}$ (for further details on Huffman trees and the adaptive
	Huffman algorithm, the reader is referred to \cite{ds:1}).
\newline\indent
	The codeword associated to the symbol $\sigma$ in the Huffman tree $\mathcal{T}_{i}(w)$ is denoted by
	${\it code}(\sigma,\mathcal{T}_{i}(w))$, for all $i\in\{0,1,\ldots,h\}$.
\begin{theorem}
\label{thm:ah}
	Adaptive Huffman encodings are particular cases of encodings by GA codes.
\end{theorem}
\begin{proof}
	Let $\Sigma$ and $\Delta$ be two alphabets, $w$ a string over $\Sigma$,
	and
	$F:\mathbb{N}^{*}\times\Sigma^{+}\rightarrow\Sigma^{*}$, $c_{F}:\Sigma\times\Sigma^{*}\rightarrow\Delta^{+}$
	two functions.
	Let us consider the function $F$ given by:
\begin{displaymath}
	F(i,\sstring{m})=
	\left\{
		\begin{array}{ll}
		\lambda								& \textrm{if $i=1$ or $i>m$.}		\\
		\sstring{i-1}						& \textrm{otherwise.}
		\end{array}
	\right.
\end{displaymath}
	and the function $c_{F}$ by $c_{F}(\sigma,u)={\it code}(\sigma,\mathcal{T}_{|u|}(u))$,
	for all $(\sigma,u)\in\Sigma\times\Sigma^{*}$.
	Let us assume that $\overline{c_{F}}$
	is not injective, that is, $\exists$ $u\sigma{v}$, $u\sigma'v'\in\Sigma^{+}$
	such that $\sigma,\sigma'\in\Sigma$, $\sigma\neq\sigma'$ and
	$\overline{c_{F}}(u\sigma{v})=\overline{c_{F}}(u\sigma'v')$.
	The previous equality can be rewritten by:
\begin{equation}
	\overline{c_{F}}(u){code}(\sigma,\mathcal{T}_{|u|}(u))\overline{c_{F}}(v)=
	\overline{c_{F}}(u){code}(\sigma',\mathcal{T}_{|u|}(u))\overline{c_{F}}(v').
\end{equation}
	Due to the prefix property of the set
	$\{{code}(\sigma,\mathcal{T}_{|u|}(u)),{code}(\sigma',\mathcal{T}_{|u|}(u))\}$, 
	the equality (3) cannot hold true, which leads to the conclusion that our assumption is false.
	Thus, we conclude that $c_{F}$ is a GA code,
	which proves the theorem.
\end{proof}
\begin{remark}
\label{rmk:3}
	If $u$ is a prefix of $w$, then $\mathcal{T}_{i}(u)=\mathcal{T}_{i}(w)$, for all $i\leq{|u|}$.
\end{remark}
\begin{example}
\label{exmp:ah}
	Let $\Sigma=\{\ttup{a},\ttup{b},\ttup{c},\ttup{d}\}$ be an alphabet, and $w=\ttup{bcabd}\in\Sigma^{+}$. Applying the adaptive
	Huffman algorithm to the input string $w$, we get the following Huffman trees.
\setlength{\unitlength}{1pt}
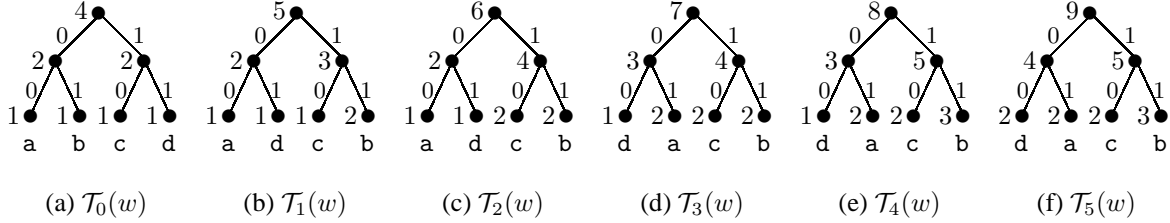
\begin{figure}[hbtp]
\begin{picture}(453,80)
\small{
%************ T0 ************
	\put(40,80){\circle*{5}}
	\put(25,79){\rotatebox{-45}{\line(0,-1){21}}}
	\put(41.6,79){\rotatebox{45}{\line(0,-1){21}}}
	\put(23.8,62){\circle*{5}}
	\put(57.2,62){\circle*{5}}
	
	\put(14.5,60.2){\rotatebox{-25}{\line(0,-1){20}}}
	\put(24.5,60.2){\rotatebox{25}{\line(0,-1){20}}}
	\put(14.4,41){\circle*{5}}
	\put(33,41){\circle*{5}}
	
	\put(48.5,60.2){\rotatebox{-25}{\line(0,-1){20}}}
	\put(58.5,60.2){\rotatebox{25}{\line(0,-1){20}}}
	\put(48.4,41){\circle*{5}}
	\put(67,41){\circle*{5}}

	\put(11.4,28){$\ttup{a}$}
	\put(30,28){$\ttup{b}$}
	\put(45.4,28){$\ttup{c}$}
	\put(64,28){$\ttup{d}$}
	
	\put(5.4,38){$1$}
	\put(24,38){$1$}
	\put(39.4,38){$1$}
	\put(58,38){$1$}
	\put(14.8,59){$2$}
	\put(48.2,59){$2$}
	\put(31,78){$4$}

	\put(12.4,48){\footnotesize{$0$}}
	\put(30,48){\footnotesize{$1$}}
	\put(46.4,48){\footnotesize{$0$}}
	\put(64,48){\footnotesize{$1$}}
	\put(24,69){\footnotesize{$0$}}
	\put(53,69){\footnotesize{$1$}}

	\put(20,5){(a) $\mathcal{T}_{0}(w)$}
%************ T1 ************
	\put(115,80){\circle*{5}}
	\put(100,79){\rotatebox{-45}{\line(0,-1){21}}}
	\put(116.6,79){\rotatebox{45}{\line(0,-1){21}}}
	\put(98.8,62){\circle*{5}}
	\put(132.2,62){\circle*{5}}
	
	\put(89.5,60.2){\rotatebox{-25}{\line(0,-1){20}}}
	\put(99.5,60.2){\rotatebox{25}{\line(0,-1){20}}}
	\put(89.4,41){\circle*{5}}
	\put(108,41){\circle*{5}}
	
	\put(123.5,60.2){\rotatebox{-25}{\line(0,-1){20}}}
	\put(133.5,60.2){\rotatebox{25}{\line(0,-1){20}}}
	\put(123.4,41){\circle*{5}}
	\put(142,41){\circle*{5}}

	\put(86.4,28){$\ttup{a}$}
	\put(105,28){$\ttup{d}$}
	\put(120.4,28){$\ttup{c}$}
	\put(139,28){$\ttup{b}$}
	
	\put(80.4,38){$1$}
	\put(99,38){$1$}
	\put(114.4,38){$1$}
	\put(133,38){$2$}
	\put(89.8,59){$2$}
	\put(123.2,59){$3$}
	\put(106,78){$5$}

	\put(87.4,48){\footnotesize{$0$}}
	\put(105,48){\footnotesize{$1$}}
	\put(121.4,48){\footnotesize{$0$}}
	\put(139,48){\footnotesize{$1$}}
	\put(99,69){\footnotesize{$0$}}
	\put(128,69){\footnotesize{$1$}}

	\put(95,5){(b) $\mathcal{T}_{1}(w)$}
%************ T2 ************
	\put(190,80){\circle*{5}}
	\put(175,79){\rotatebox{-45}{\line(0,-1){21}}}
	\put(191.6,79){\rotatebox{45}{\line(0,-1){21}}}
	\put(173.8,62){\circle*{5}}
	\put(207.2,62){\circle*{5}}
	
	\put(164.5,60.2){\rotatebox{-25}{\line(0,-1){20}}}
	\put(174.5,60.2){\rotatebox{25}{\line(0,-1){20}}}
	\put(164.4,41){\circle*{5}}
	\put(183,41){\circle*{5}}
	
	\put(198.5,60.2){\rotatebox{-25}{\line(0,-1){20}}}
	\put(208.5,60.2){\rotatebox{25}{\line(0,-1){20}}}
	\put(198.4,41){\circle*{5}}
	\put(217,41){\circle*{5}}

	\put(161.4,28){$\ttup{a}$}
	\put(180,28){$\ttup{d}$}
	\put(195.4,28){$\ttup{c}$}
	\put(214,28){$\ttup{b}$}
	
	\put(155.4,38){$1$}
	\put(174,38){$1$}
	\put(189.4,38){$2$}
	\put(208,38){$2$}
	\put(164.8,59){$2$}
	\put(198.2,59){$4$}
	\put(181,78){$6$}

	\put(162.4,48){\footnotesize{$0$}}
	\put(180,48){\footnotesize{$1$}}
	\put(196.4,48){\footnotesize{$0$}}
	\put(214,48){\footnotesize{$1$}}
	\put(174,69){\footnotesize{$0$}}
	\put(203,69){\footnotesize{$1$}}

	\put(170,5){(c) $\mathcal{T}_{2}(w)$}
%************ T3 ************
	\put(265,80){\circle*{5}}
	\put(250,79){\rotatebox{-45}{\line(0,-1){21}}}
	\put(266.6,79){\rotatebox{45}{\line(0,-1){21}}}
	\put(248.8,62){\circle*{5}}
	\put(282.2,62){\circle*{5}}
	
	\put(239.5,60.2){\rotatebox{-25}{\line(0,-1){20}}}
	\put(249.5,60.2){\rotatebox{25}{\line(0,-1){20}}}
	\put(239.4,41){\circle*{5}}
	\put(258,41){\circle*{5}}
	
	\put(273.5,60.2){\rotatebox{-25}{\line(0,-1){20}}}
	\put(283.5,60.2){\rotatebox{25}{\line(0,-1){20}}}
	\put(273.4,41){\circle*{5}}
	\put(292,41){\circle*{5}}

	\put(236.4,28){$\ttup{d}$}
	\put(255,28){$\ttup{a}$}
	\put(270.4,28){$\ttup{c}$}
	\put(289,28){$\ttup{b}$}
	
	\put(230.4,38){$1$}
	\put(249,38){$2$}
	\put(264.4,38){$2$}
	\put(283,38){$2$}
	\put(239.8,59){$3$}
	\put(273.2,59){$4$}
	\put(256,78){$7$}

	\put(237.4,48){\footnotesize{$0$}}
	\put(255,48){\footnotesize{$1$}}
	\put(271.4,48){\footnotesize{$0$}}
	\put(289,48){\footnotesize{$1$}}
	\put(249,69){\footnotesize{$0$}}
	\put(278,69){\footnotesize{$1$}}

	\put(245,5){(d) $\mathcal{T}_{3}(w)$}
%************ T4 ************
	\put(340,80){\circle*{5}}
	\put(325,79){\rotatebox{-45}{\line(0,-1){21}}}
	\put(341.6,79){\rotatebox{45}{\line(0,-1){21}}}
	\put(323.8,62){\circle*{5}}
	\put(357.2,62){\circle*{5}}
	
	\put(314.5,60.2){\rotatebox{-25}{\line(0,-1){20}}}
	\put(324.5,60.2){\rotatebox{25}{\line(0,-1){20}}}
	\put(314.4,41){\circle*{5}}
	\put(333,41){\circle*{5}}
	
	\put(348.5,60.2){\rotatebox{-25}{\line(0,-1){20}}}
	\put(358.5,60.2){\rotatebox{25}{\line(0,-1){20}}}
	\put(348.4,41){\circle*{5}}
	\put(367,41){\circle*{5}}

	\put(311.4,28){$\ttup{d}$}
	\put(330,28){$\ttup{a}$}
	\put(345.4,28){$\ttup{c}$}
	\put(364,28){$\ttup{b}$}
	
	\put(305.4,38){$1$}
	\put(324,38){$2$}
	\put(339.4,38){$2$}
	\put(358,38){$3$}
	\put(314.8,59){$3$}
	\put(348.2,59){$5$}
	\put(331,78){$8$}

	\put(312.4,48){\footnotesize{$0$}}
	\put(330,48){\footnotesize{$1$}}
	\put(346.4,48){\footnotesize{$0$}}
	\put(364,48){\footnotesize{$1$}}
	\put(324,69){\footnotesize{$0$}}
	\put(353,69){\footnotesize{$1$}}

	\put(320,5){(e) $\mathcal{T}_{4}(w)$}
%************ T5 ************
	\put(415,80){\circle*{5}}
	\put(400,79){\rotatebox{-45}{\line(0,-1){21}}}
	\put(416.6,79){\rotatebox{45}{\line(0,-1){21}}}
	\put(398.8,62){\circle*{5}}
	\put(432.2,62){\circle*{5}}
	
	\put(389.5,60.2){\rotatebox{-25}{\line(0,-1){20}}}
	\put(399.5,60.2){\rotatebox{25}{\line(0,-1){20}}}
	\put(389.4,41){\circle*{5}}
	\put(408,41){\circle*{5}}
	
	\put(423.5,60.2){\rotatebox{-25}{\line(0,-1){20}}}
	\put(433.5,60.2){\rotatebox{25}{\line(0,-1){20}}}
	\put(423.4,41){\circle*{5}}
	\put(442,41){\circle*{5}}

	\put(386.4,28){$\ttup{d}$}
	\put(405,28){$\ttup{a}$}
	\put(420.4,28){$\ttup{c}$}
	\put(439,28){$\ttup{b}$}
	
	\put(380.4,38){$2$}
	\put(399,38){$2$}
	\put(414.4,38){$2$}
	\put(433,38){$3$}
	\put(389.8,59){$4$}
	\put(423.2,59){$5$}
	\put(406,78){$9$}

	\put(387.4,48){\footnotesize{$0$}}
	\put(405,48){\footnotesize{$1$}}
	\put(421.4,48){\footnotesize{$0$}}
	\put(439,48){\footnotesize{$1$}}
	\put(399,69){\footnotesize{$0$}}
	\put(428,69){\footnotesize{$1$}}

	\put(395,5){(f) $\mathcal{T}_{5}(w)$}
}
\end{picture}
\caption{The Huffman trees associated to $w$: $\mathcal{T}_{0}(w)$, $\mathcal{T}_{1}(w)$,
		$\mathcal{T}_{2}(w)$, $\mathcal{T}_{3}(w)$, $\mathcal{T}_{4}(w)$, and $\mathcal{T}_{5}(w)$.}
\end{figure}
\hspace{0pt}\newline
	Let $F:\mathbb{N}^{*}\times\Sigma^{+}\rightarrow\Sigma^{*}$,
	$c_{F}:\Sigma\times\Sigma^{*}\rightarrow\{0,1\}^{+}$ be constructed as above.
	Then, we encode $w$ by:
	$\overline{c_{F}}(\ttup{bcabd})=c_{F}(\ttup{b},\lambda)c_{F}(\ttup{c},\ttup{b})c_{F}(\ttup{a},\ttup{bc})c_{F}(\ttup{b},\ttup{bca})c_{F}(\ttup{d},\ttup{bcab})$
	$={\it code}(\ttup{b},\mathcal{T}_{0}(\lambda)){\it code}(\ttup{c},\mathcal{T}_{1}(\ttup{b}))$
	${\it code}(\ttup{a},\mathcal{T}_{2}(\ttup{bc})){\it code}(\ttup{b},\mathcal{T}_{3}(\ttup{bca})){\it code}(\ttup{d},\mathcal{T}_{4}(\ttup{bcab}))=0110001100$.
\end{example}

\section{GA codes and Lempel-Ziv codes}
	The aim of this section is to prove that Lempel-Ziv encodings are particular cases of encodings
	by GA codes. Let $\Sigma$ and $\Delta$ be two alphabets such that $\{0,1,\ldots,9\}\cap\Sigma=\emptyset$.
	First, we recall the Lempel-Ziv parsing procedure of an input data string $w$,
	where $w=w_{1}w_{2}\ldots{w_{h}}$ is a string over $\Sigma$.
	For more details, the reader is referred to \cite{zl:2,zl:1}.
\newline\indent
	The first variable-length block arising from the Lempel-Ziv parsing of the data string $w$ is $w_{1}$.
	The second block in the parsing is the shortest prefix of $w_{2}\ldots{w_{h}}$ which is not
	equal to $w_{1}$. Consider that this second block is $w_{2}\ldots{w_{j}}$.
	Then, the third block will be the shortest prefix of $w_{j+1}\ldots{w_{h}}$ which is not equal
	to either $w_{1}$ or $w_{2}\ldots{w_{j}}$.
	Suppose the Lempel-Ziv parsing
	has produced the first $k$ variable-length blocks $B_{1},B_{2},\ldots,{B_{k}}$ in the parsing,
	and $w^{(k)}$ is that part left of $w$ after $B_{1},B_{2},\ldots,{B_{k}}$ have been removed.
	Then, the next block $B_{k+1}$ in the parsing is the shortest prefix of $w^{(k)}$ which
	is not equal to any of the preceding blocks $B_{1},B_{2},\ldots,{B_{k}}$ (if there is no such
	block, then $B_{k+1}=w^{(k)}$ and the Lempel-Ziv parsing procedure terminates).
\begin{theorem}
\label{thm:lz}
	Lempel-Ziv encodings are particular cases of encodings by GA codes.
\end{theorem}
\begin{proof}
	Let $\Sigma_{1}=\Sigma\cup\{0,1,\ldots,9\}$ be an alphabet, 
	$\sigma_{f}\in\Sigma$ a fixed symbol, and let
	$F:\mathbb{N}^{*}\times\Sigma_{1}^{+}\rightarrow\Sigma_{1}^{*}$,
	$c_{F}:\Sigma_{1}\times\Sigma_{1}^{*}\rightarrow\{0,1\}^{*}$
	be two functions.
\newline\indent
	Let us consider $F$ given by $F(i,\sstring{m})=i_{1}i_{2}\ldots{i_{q}}\sigma_{f}\sstring{m}$,
	for all $i\in{\mathbb{N}^{*}}$ and $\sstring{m}\in\Sigma_{1}^{+}$, where
	$i_{1},i_{2},\ldots,i_{q}\in\{0,1,\ldots,9\}$ are the digits corresponding to $i$ (from left to right).
\newline\indent
	Let $u=u_{1}u_{2}\ldots{u_{p}}$ be a string over $\Sigma_{1}$, that is, $u_{i}\in\Sigma_{1}$ for all $i\in\{1,2,\ldots,p\}$.
	Consider the following notations.
\begin{itemize}
\item	${\it fixed}(u)=
	\left\{
	\begin{array}{ll}
	1				& \textrm{if $p\geq{3}$ and 
						$\exists$ $i\in\{2,3,\ldots,p-1\}$, such that $u_{i}=\sigma_{f}$}\\
					& \textrm{and $u_{j}\in\{0,1,\ldots,9\}$ for all $j\in\{1,2,\ldots,i-1\}$.}\\
	0				& \textrm{otherwise.}
	\end{array}
	\right.$
\item ${\it left}(u)=
	\left\{
	\begin{array}{ll}
	u_{1}u_{2}\ldots{u_{r}} 	& \textrm{if ${\it fixed}(u)=1$, $u_{i}\in\{0,1,\ldots,9\}$ for all}\\
								& \textrm{$i\in\{1,2,\ldots,r\}$, and $u_{r+1}=\sigma_{f}$.}\\
	\lambda						& \textrm{otherwise.}
	\end{array}
	\right.$
\item ${\it right}(u)=
	\left\{
	\begin{array}{ll}
	v				 	& \textrm{if ${\it fixed}(u)=1$ and $u={\it left}(u)\sigma_{f}v$.}\\
	\lambda				& \textrm{otherwise.}
	\end{array}
	\right.$
\item ${\it goodpos}(u)=
	\left\{
	\begin{array}{ll}
	1				 	& \textrm{if ${\it fixed}(u)=1$ and
							$|{\it left}(u)|+2\leq{{\it left}(u)}\leq{|u|}$.}\\
	0					& \textrm{otherwise.}
	\end{array}
	\right.$
\end{itemize}
	Let us consider $c_{F}$ given by
\begin{displaymath}
	c_{F}(\sigma,\sstring{m})=
	\left\{
		\begin{array}{ll}
		{\it LZ}(\sigma,\sstring{m})	& \textrm{if ${\it fixed}(\sstring{m})=1$, ${\it goodpos}(\sstring{m})=1$}\\
							%& \textrm{${goodpos}(\sstring{m})=1$}\\
							& \textrm{and $\sigma=\sigma_{{left}(\sstring{m})}$.}\\
		\lambda				& \textrm{otherwise.}
	\end{array}
	\right.
\end{displaymath}
	where ${\it LZ}(\sigma,\sstring{m})$ is defined as follows:
	let $B_{1},B_{2},\ldots,B_{t}$ be the blocks arising
	from the Lempel-Ziv parsing of the string ${\it right}(\sstring{m})$, and
\begin{equation}
	B_{z}=\sigma_{|{\it left}(\sstring{m})|+2+j_{1}}\ldots\sigma_{|{\it left}(\sstring{m})|+2+j_{2}},
\end{equation}
	where $z\in\{1,\ldots,t\}$, $0\leq{j_{1}}\leq{j_{2}}\leq{|{\it rigth}(\sstring{m})|-1}$, and
\begin{equation}
|{\it left}(\sstring{m})|+2+j_{1}\leq{{\it left}(\sstring{m})}\leq{|{\it left}(\sstring{m})|+2+j_{2}}.
\end{equation}
	If ${\it left}(\sstring{m})=|{\it left}(\sstring{m})|+2+j_{2}$, then let
	${\it LZ}(\sigma,\sstring{m})$ be
	the codeword associated by the Lempel-Ziv data compression algorithm to the block $B_{z}$. Otherwise,
	we consider that ${\it LZ}(\sigma,\sstring{m})=\lambda$. One can easily verify that
	$\overline{c_{F}}(\sstring{m})$ is the encoding of $\sstring{m}$ by the Lempel-Ziv data compression
	algorithm, for all $\sstring{m}\in\Sigma_{1}^{+}$.
	Thus, we have obtained that $\overline{c_{F}}$ is injective, which proves the theorem.
\end{proof}
\begin{example}
\label{exmp:lz}
	Let $\Sigma=\{\ttup{a},\ttup{b},\ttup{c}\}$, $\Sigma_{1}=\Sigma\cup\{0,1,\ldots,9\}$ be two alphabets,
	and let $F:\mathbb{N}^{*}\times\Sigma_{1}^{+}\rightarrow\Sigma_{1}^{*}$,
	$c_{F}:\Sigma_{1}\times\Sigma_{1}^{*}\rightarrow\{0,1\}^{*}$ be two functions given as in Theorem \ref{thm:lz}
	(considering $\sigma_{f}=\ttup{a}$). Also, let $w=\ttup{bcc}7\ttup{ba}\in\Sigma_{1}^{+}$ be an input string.

	Applying the Lempel-Ziv parsing procedure to the input string $w$, we get the following blocks:
	$B_{1}=\ttup{b}$, $B_{2}=\ttup{c}$, $B_{3}=\ttup{c}7$, and $B_{4}=\ttup{ba}$. Let us denote by ${\it codeLZ}(B_{i})$ the
	codeword associated by the Lempel-Ziv encoder to the block $B_{i}$, for all $i\in\{1,2,3,4\}$.
	One can verify that we get the following results:
\begin{itemize}
\item ${\it codeLZ}(B_{1})=1011$,
\item ${\it codeLZ}(B_{2})=01100$,
\item ${\it codeLZ}(B_{3})=100001$,
\item ${\it codeLZ}(B_{4})=010111$.
\end{itemize}
	Finally, we encode $w=\ttup{bcc}7\ttup{ba}$ by the GA code $c_{F}$ as shown below.
\newline\indent
	$\overline{c_{F}}(w)=c_{F}(\ttup{b},F(1,\ttup{bcc}7\ttup{ba}))c_{F}(\ttup{c},F(2,\ttup{bcc}7\ttup{ba}))c_{F}(\ttup{c},F(3,\ttup{bcc}7\ttup{ba}))$
\newline\indent\hspace{40pt}
	$c_{F}(7,F(4,\ttup{bcc}7\ttup{ba}))c_{F}(\ttup{b},F(5,\ttup{bcc}7\ttup{ba}))c_{F}(\ttup{a},F(6,\ttup{bcc}7\ttup{ba}))$
\newline\indent\hspace{28.5pt}
	$=c_{F}(\ttup{b},1\ttup{abcc}7\ttup{ba})c_{F}(\ttup{c},2\ttup{abcc}7\ttup{ba})c_{F}(\ttup{c},3\ttup{abcc}7\ttup{ba})$
\newline\indent\hspace{40pt}
	$c_{F}(7,4\ttup{abcc}7\ttup{ba})c_{F}(\ttup{b},5\ttup{abcc}7\ttup{ba})c_{F}(\ttup{a},6\ttup{abcc}7\ttup{ba})$
\newline\indent\hspace{28.5pt}
	$={\it LZ}(\ttup{b},1\ttup{abcc}7\ttup{ba})\cdot{\it LZ}(\ttup{c},2\ttup{abcc}7\ttup{ba})\cdot\lambda\cdot$
	${\it LZ}(7,4\ttup{abcc}7\ttup{ba})\cdot\lambda\cdot{\it LZ}(\ttup{a},6\ttup{abcc}7\ttup{ba})$
\newline\indent\hspace{28.5pt}
	$={\it codeLZ}(B_{1}){\it codeLZ}(B_{2}){\it codeLZ}(B_{3}){\it codeLZ}(B_{4})$
\newline\indent\hspace{28.5pt}
	$=101101100100001010111$.
\end{example}

\section{Adaptive codes and convolutional codes}
	Convolutional codes \cite{p:h:1} are one of the most widely used channel codes in practical communication systems.
	These codes are developed with a separate strong mathematical structure and are primarily used for real time
	error correction. Convolutional codes convert the entire data stream into one single codeword: the encoded bits depend
	not only on the current $k$ input bits, but also on past input bits.
	The same strategy is used by adaptive variable-length codes. The aim of this section is to discuss the connection
	between adaptive codes and convolutional codes. Specifically, we show how a convolutional code can be modelled as an adaptive code.
	Before stating the results, let us first present a brief description of convolutional codes.
	
	Convolutional codes are commonly specified by three parameters: $n$, $k$, and $m$, where
\begin{itemize}
\item $n$ is the number of output bits,
\item $k$ is the number of input bits,
\item and $m$ is the number of memory registers.
\end{itemize}
	The quantity $km$ is called the {\it constraint length}, and represents the number of bits in the encoder memory that affect
	the generation of the $n$ output bits.
	Also, the quantity $k/n$ is called the {\it code rate}, and is a measure of the efficiency of the code.
	A convolutional code with parameters $n$, $k$, $m$ is usually referred to as an $(n,k,m)$ convolutional code.
	For an $(n,k,m)$ convolutional code, the encoding procedure is entirely defined by $n$ {\it generator polynomials}.
	Usually, these generator polynomials are represented as binary $(m+1)$-tuples.
	Also, throughout this section, we consider only $(n,1,m)$ convolutional codes.
	
	Let us consider an $(n,1,m)$ convolutional code with $P_{1},P_{2},\ldots,P_{n}$ being its generator polynomials, and let
	$x=x_{1}x_{2}\ldots{x_{t}}\in\{0,1\}^{+}$ be an input data string.
	The string $x$ is encoded by $y=y_{1}y_{2}\ldots{y_{nt}}$, where the substring $y_{in+1}\ldots{y_{in+n}}$ encodes
	the input bit $x_{i+1}$, for all $i\in\{0,1,\ldots,t-1\}$.
	Precisely, if $i\in\{0,1,\ldots,t-1\}$ and $j\in\{1,2,\ldots,n\}$, then
\begin{center}
	$y_{in+j}=w_{i-i_{1}^{j}+1}\oplus{w_{i-i_{2}^{j}+1}}\oplus\ldots\oplus{w_{i-i_{q(j)}^{j}+1}}$,
\end{center}
	where $\{i_{1}^{j},i_{2}^{j},\ldots,i_{q(j)}^{j}\}=\{z\in\{1,2,\ldots,m+1\} \mid P_{j}.z=1\}$,
	$i_{1}^{j}\leq{i_{2}^{j}}\leq\ldots\leq{i_{q(j)}^{j}}$, $\oplus$ denotes the modulo-2 addition, and
\begin{displaymath}
	w_{i-l}=
	\left\{
		\begin{array}{ll}
		x_{i-l+1}					& \textrm{if $i-l+1\geq{1}$.}\\
		0							& \textrm{otherwise.}
	\end{array}
	\right.
\end{displaymath}
	for all $l\in\{0,1,\ldots,m\}$.
\begin{example}
\label{exmp:conv}
	Let us consider a $(2,1,2)$ convolutional code with $P_{1}=(0,1,1),P_{2}=(1,0,1)$ being its generator polynomials.
	This convolutional code can be represented graphically as in the figure below.
\begin{figure}[hbtp]
\setlength{\unitlength}{1pt}
\begin{picture}(400,70)(-40,0)
	\put(50,40){input}
	\put(75,42){\vector(1,0){40}}
	\put(115,32){\framebox(20,20){$m_{1}$}}
	\put(135,42){\vector(1,0){40}}
	\put(175,32){\framebox(20,20){$m_{2}$}}
	\put(195,42){\line(1,0){40}}
	\put(229.5,60.5){\Large{$\oplus$}}
	\put(229.5,16.5){\Large{$\oplus$}}

	\put(235,42){\vector(0,1){17.5}}
	\put(235,42){\vector(0,-1){17}}
	\put(155,42){\line(0,1){22}}
	\put(155,64){\vector(1,0){75.2}}
	\put(95,42){\line(0,-1){22}}
	\put(95,20){\vector(1,0){135.2}}
	\put(239,64){\vector(1,0){40}}
	\put(239,20){\vector(1,0){40}}
	\put(282,62){output 1}
	\put(282,18){output 2}
\end{picture}
\caption{A $(2,1,2)$ convolutional code.}
\end{figure}
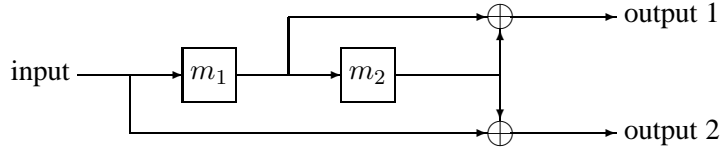
\hspace{0pt}
	Let us now describe the encoding mechanism. Let $b$ be the current input bit being encoded, and let $b_{1}$ and $b_{2}$ be the
	current bits stored in the memory registers $m_{1}$ and $m_{2}$, respectively.
	Given that $P_{1}=(0,1,1)$, the first output bit is obtained by adding (modulo-2) $b_{1}$ and $b_{2}$.
	The second bit is obtained by adding (modulo-2) $b$ and $b_{2}$.
	After both output bits have been obtained, $b$ and $b_{1}$ become the new values stored in the memory registers $m_{1}$ and $m_{2}$,
	respectively.
	For example, if $x=0101$ is an input bitstring, one can verify that the output is $00011010$ (for each input bit, the output
	is obtained by concatenating the two output bits).
\end{example}
\begin{theorem}
\label{thm:conv}
	Any $(n,1,m)$ convolutional code with $P_{1},P_{2},\ldots,P_{n}$ being its generator polynomials, and satisfying
	the condition
\begin{center}
	$\{z\in\{1,2,\ldots,n\} \mid P_{z}.1=1\}\neq\emptyset$,
\end{center}
	is an adaptive code of order $m$.
\end{theorem}
\begin{proof}
	Let $c:\{0,1\}\times\{0,1\}^{\leq{m}}\rightarrow\{0,1\}^{n}$ be a function.
	Consider an $(n,1,m)$ convolutional code with $P_{1},P_{2},\ldots,P_{n}$ being its generator polynomials.
	Also, let us consider that $c$ is given by:
\begin{center}
	$c(x,x_{1}x_{2}\ldots{x_{p}})=P_{1}[xx_{p}x_{p-1}\ldots{x_{1}}z_{p}^{m}]P_{2}[xx_{p}x_{p-1}\ldots{x_{1}}z_{p}^{m}]\ldots$
	$P_{n}[xx_{p}x_{p-1}\ldots{x_{1}}z_{p}^{m}]$,
\end{center}
	for all $x\in\{0,1\}$ and $x_{1}x_{2}\ldots{x_{p}}\in\{0,1\}^{\leq{m}}$, where
\begin{itemize}
\item $z_{p}^{m}=\underbrace{00\ldots{0}}_{m-p}$,
\item and $P_{j}[b_{1}b_{2}\ldots{b_{m+1}}]=b_{i_{1}^{j}}\oplus{b_{i_{2}^{j}}}\oplus\ldots\oplus{b_{i_{q(j)}^{j}}}$,
	with $\{i_{1}^{j},i_{2}^{j},\ldots,i_{q(j)}^{j}\}=\{z \mid P_{j}.z=1\}$ and
	$i_{1}^{j}\leq{i_{2}^{j}}\leq\ldots\leq{i_{q(j)}^{j}}$.
\end{itemize}
	Let $b_{1}b_{2}\ldots{b_{q}}\in\{0,1\}^{\leq{m}}$. By hypothesis, there exists $j\in\{1,2,\ldots,n\}$ such that
	$P_{j}.1=1$. This leads to the conclusion that
\begin{center}
	$\{c(0,b_{1}b_{2}\ldots{b_{q}}),c(1,b_{1}b_{2}\ldots{b_{q}})\}$
\end{center}
	is a prefix code.
	Thus, we have obtained that $C_{u}$ (as defined in section 2) is a prefix code, for all $u\in\{0,1\}^{\leq{m}}$.
	According to Theorem \ref{thm:1}, $c$ is an adaptive code of order $m$.
\end{proof}
\section{A cryptographic scheme based on convolutional codes}
	The results presented in the previous section lead to an efficient data encryption scheme.
	Specifically, every $(1,1,m)$ convolutional code satisfying the condition in Theorem \ref{thm:conv} can be used
	for data encryption (and decryption), without any additional information. Let us consider an $(1,1,m)$ convolutional code
	with $P$ being its generator polynomial. If $P.1=0$ (that is, the condition in Theorem \ref{thm:conv} is not satisfied), then
	the output bits depend {\it only} on the bits stored in the memory registers.
	For example, let $b$ be the current input bit, and $b_{1},b_{2},\ldots,b_{m}$ the bits stored in the memory registers before encoding
	the bit $b$. The output bit $b_{{\it out}}$ depends, in this case, only on the bits $b_{1},b_{2},\ldots,b_{m}$. 
	This makes the decryption procedure impossible (without
	any additional information), since the output cannot be uniquely decoded.
	Thus, we consider only $(1,1,m)$ convolutional codes that satisfy the condition given in Theorem \ref{thm:conv}.
	Also, we consider that any $(1,1,m)$ convolutional code is completely specified by
\begin{itemize}
\item $P$, its generator polynomial,
\item and a binary $m$-uple $Q$, where $Q.i$ denotes the bit stored initially in the memory register $m_{i}$, for all $i\in\{1,2,\ldots,m\}$.
\end{itemize}
	{\bf Public and Private Keys.}
	Let us denote by ${\it Public}$ the set of public keys, and by ${\it Private}$ the set of private keys. There are three parameters
	in our cryptographic scheme: $m$, $P$, and $Q$. Note that by making $P$ and/or $Q$ available to any user, the parameter $m$ is
	implicitly made available as well (since $P$ consists of $m+1$ elements, and $Q$ has $m$ elements).
	Thus, if $P$ and $Q$ are both public keys, then the information can be correctly decoded.
	Except for the case when both $P$ and $Q$ are public keys, all other cases lead to a powerful cryptographic scheme.
	The parameters $P$ and $Q$ shouldn't normally be among the public keys, since both $P$ and $Q$ give partial information about
	the encryption/decryption procedures. Thus, we consider that in practice only the parameter $m$ should be included among the public keys.
	Keeping all three parameters as private keys increases the security level as well (by a constant factor).
\begin{table}[htbp]
\caption{Possible ways of partitioning the keys.}
\begin{center}
\begin{tabular}{|c|c|c|c|}
\hline
${\it Public}$	& ${\it Private}$	& Security level 	& Complexity 					\\ \hline
$\emptyset$		& $\{P,Q,m\}$		& High				& $\mathcal{O}(4^{m})$			\\ 
$\{m\}$			& $\{P,Q\}$			& High				& $\mathcal{O}(4^{m})$			\\ 
$\{P\}$			& $\{Q\}$			& High				& $\mathcal{O}(2^{m})$			\\ 
$\{Q\}$			& $\{P\}$			& High				& $\mathcal{O}(2^{m})$			\\ \hline
\end{tabular}
\end{center}
\end{table}
\hspace{0pt}
\newline
	{\bf Security and Complexity.}
	There are four ways of partitioning the keys, as shown in the table above. Note that if $P$ or $Q$ is a public key, then
	it doesn't make sense to include $m$ as a public or private key, since if $P$ or $Q$ is made available then $m$ is implicitly
	a public key. Let us discuss each case separately.
\begin{description}
\item[${\it Public}=\emptyset$ and ${\it Private}=\{P,Q,m\}$.]
	In this case, an unauthorized user has no information about the encryption/decryption procedure.
	A possible attack cannot be more efficient than a naive search, starting with $m=1$ and trying all possible cases for
	$P$ and $Q$.
	Since there are $2^{m}$ possible binary $m$-tuples, we can conclude that the total number of decoding attempts
	is at most
\begin{displaymath}
	2^{1}\cdot{2^{1}}+2^{2}\cdot{2^{2}}+\ldots{+}2^{m}\cdot{2^{m}}=\frac{4^{m+1}-4}{3}=\frac{2^{2m+2}-4}{3}.
\end{displaymath}
	For example, if $m=100$, then the total number of decoding attempts is at most
\begin{displaymath}
	\frac{2^{202}-4}{3}\approx{2.1\cdot{10^{60}}}.
\end{displaymath}
	Definitely, the scheme is highly efficient in this case.
\item[${\it Public}=\{m\}$ and ${\it Private}=\{P,Q\}$.]
	Even if $m$ is a public key, the efficiency of our scheme is not affected at all.
	A possible attack must try all possible cases for $P$ and $Q$ (in the worst case).
	Thus, the total number of decoding attempts is at most $2^{m}\cdot{2^{m}}=2^{2m}$.
	For $m=100$, the total number of decoding attempts is at most $2^{200}\approx{1.6\cdot{10^{60}}}$.
\item[${\it Public}=\{P\}$ and ${\it Private}=\{Q\}$.]
	Since only $Q$ is a private key in this case, we can conclude that the total number of decoding attempts is at most
	$2^{m}$. For $m=100$, $2^{100}\approx{1.2\cdot{10^{30}}}$.
\item[${\it Public}=\{Q\}$ and ${\it Private}=\{P\}$.]
	The total number of decoding attempts is at most $2^{m}$, since only $P$ is a private key in this case.
\end{description}
	{\bf Encryption and Decryption.}
	A detailed description of the encryption algorithm is provided below. Note that $m$ is a positive integer,
	$P$ is a binary $(m+1)$-tuple that satisfies the condition in Theorem \ref{thm:conv}, and $Q$ is a binary $m$-tuple. 
\begin{figure}[ht]
\begin{center}
{\footnotesize
	\fbox{
		\begin{minipage}{300pt}
		\begin{tabbing}
		\hspace*{5mm}\=\hspace{5mm}\=\hspace{5mm}\=\hspace{5mm}\=\hspace{5mm}\=\hspace{5mm}\=  \kill
		{\tt Input:} $m$, $P$, $Q$, and $x=x_{1}x_{2}\ldots{x_{t}}\in\{0,1\}^{+}$\\
		{\tt Output:} $y\in\{0,1\}^{+}$\\
		\rule[3pt]{1.0\textwidth}{0.3pt}\\
		$S\leftarrow\emptyset$; $y\leftarrow\lambda$\\
		{\tt For} $i\leftarrow{2}$ to $m+1$ {\tt do}\\
		\>\>{\tt If} $P.i=1$ {\tt then}\\
		\>\>\> $S\leftarrow{S\cup\{i-1\}}$\\
		\>\>{\tt Endif}\\
		{\tt Endfor}\\
		{\tt For} $i\leftarrow{1}$ to $t$ {\tt do}\\
		\>\> $z\leftarrow{x_{i}}$\\
		\>\>{\tt For each} $j\in{S}$ {\tt do}\\
		\>\>\> $z\leftarrow{z\oplus{Q.j}}$\\
		\>\>{\tt Endfor}\\
		\>\> $y\leftarrow{y\cdot{z}}$; $Q\leftarrow{Q\vartriangleright{m}}$; $Q\leftarrow{(x_{i})\vartriangle{Q}}$\\
		{\tt Endfor}
		\end{tabbing}
		\end{minipage}
	}
}
\end{center}
\caption{Convolutional encryption.}
\end{figure}
\hspace{0pt}
\newline
	As mentioned in the beginning of this section, the decryption algorithm is based on the equality $P.1=1$.
	Let $y_{i}$ be the current bit being decoded, $Q$ the content of the memory registers before decoding $y_{i}$,
	$S=\{i_{1},i_{2},\ldots,i_{j}\}$ the set of indexes of those memory registers that contribute to the output bit,
	and $z=Q.i_{1}\oplus{Q.i_{2}}\oplus\ldots\oplus{Q.i_{j}}$.
	If $y_{i}=0$, we can conclude that $x_{i}=z$ (since $x_{i}\oplus{z}=0$).
	Otherwise, if $y_{i}=1$, it follows that $x_{i}=\overline{z}$, where $\overline{z}$ denotes the complement of $z$.
	A complete description of the algorithm is given below.
\begin{figure}[ht]
\begin{center}
{\footnotesize
	\fbox{
		\begin{minipage}{300pt}
		\begin{tabbing}
		\hspace*{5mm}\=\hspace{5mm}\=\hspace{5mm}\=\hspace{5mm}\=\hspace{5mm}\=\hspace{5mm}\=  \kill
		{\tt Input:} $y=y_{1}y_{2}\ldots{y_{t}}\in\{0,1\}^{+}$, $m$, $P$, and $Q$\\
		{\tt Output:} $x\in\{0,1\}^{+}$\\
		\rule[3pt]{1.0\textwidth}{0.3pt}\\
		$S\leftarrow\emptyset$; $x\leftarrow\lambda$\\
		{\tt For} $i\leftarrow{2}$ to $m+1$ {\tt do}\\
		\>\>{\tt If} $P.i=1$ {\tt then}\\
		\>\>\> $S\leftarrow{S\cup\{i-1\}}$\\
		\>\>{\tt Endif}\\
		{\tt Endfor}\\
		{\tt For} $i\leftarrow{1}$ to $t$ {\tt do}\\
		\>\> $z\leftarrow{0}$\\
		\>\>{\tt For each} $j\in{S}$ {\tt do}\\
		\>\>\> $z\leftarrow{z\oplus{Q.j}}$\\
		\>\>{\tt Endfor}\\
		\>\>{\tt If} $z\neq{y_{i}}$ {\tt then}\\
		\>\>\> $z\leftarrow{1}$\\
		\>\>{\tt Endif}\\
		\>\> $x\leftarrow{x\cdot{z}}$; $Q\leftarrow{Q\vartriangleright{m}}$; $Q\leftarrow{(z)\vartriangle{Q}}$\\
		{\tt Endfor}
		\end{tabbing}
		\end{minipage}
	}
}
\end{center}
\caption{Convolutional decryption.}
\end{figure}
\begin{example}
	Consider an $(1,1,2)$ convolutional code with $P=(1,0,1)$ and $Q=(0,1)$.
	Graphically, this code is represented as in the figure below.
\begin{figure}[hbtp]
\setlength{\unitlength}{1pt}
\begin{picture}(400,60)(-40,20)
	\put(50,40){input}
	\put(75,42){\vector(1,0){40}}
	\put(115,32){\framebox(20,20){$m_{1}$}}
	\put(135,42){\vector(1,0){40}}
	\put(175,32){\framebox(20,20){$m_{2}$}}
	\put(195,42){\vector(1,0){40.2}}
	\put(234.5,38.5){\Large{$\oplus$}}

	\put(95,42){\line(0,1){22}}
	\put(95,64){\line(1,0){145.2}}
	\put(240.2,64){\vector(0,-1){17}}
	\put(245,42){\vector(1,0){40}}
	\put(287,40){output}
\end{picture}
\caption{An $(1,1,2)$ convolutional code.}
\end{figure}
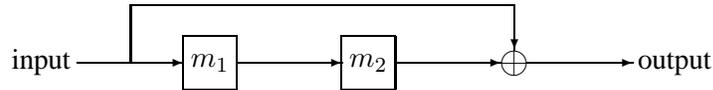
\hspace{0pt}
	Initially, the memory register $m_{1}$ stores the bit $0$ ($=Q.1$), and the memory register $m_{2}$ stores the bit $1$ ($=Q.2$).
	Let $x=001\in\{0,1\}^{+}$ be an input bitstring. Using the encryption algorithm, we encode $x$ by $y=101$.
	Given that $P.1=1$, we can use the convolutional decryption algorithm to decode $y$ into $x$ (using the private keys $m$, $P$, and $Q$).
\end{example}
\section{$(p,q)$-adaptive codes}
	In order to have more flexibility when developing applications based on adaptive codes, we introduce a natural generalization
	of adaptive codes, called $(p,q)$-adaptive codes. For example, extending the algorithms presented in \cite{t:2}
	to $(p,q)$-adaptive codes is expected to give better results. Let us give a formal definition.
\begin{definition}
\label{def:pqac}
	Let $\Sigma$ and $\Delta$ be alphabets. A function
	$c:\Sigma^{q}\times\Sigma^{\leq{p}}\rightarrow\Delta^{+}$ is called $(p,q)$-\emph{adaptive code} if its unique
	homomorphic extension $\overline{c}:\Sigma^{*}\rightarrow\Delta^{*}$, given by:
\begin{itemize}
\item $\overline{c}(\lambda)=\lambda$,
\item $\overline{c}(\sstring{m})=$
	$c(\sigma_{1}\ldots{\sigma_{q}},\lambda)$
	$c(\sigma_{2}\ldots{\sigma_{q+1}},\sigma_{1})$
	$\ldots$
	$c(\sigma_{p+1}\ldots{\sigma_{p+q}},\sigma_{1}\ldots{\sigma_{p}})$
	\newline
	$c(\sigma_{p+2}\ldots{\sigma_{p+q+1}},\sigma_{2}\ldots\sigma_{p+1})$
	$\ldots$
	$c(\sigma_{m-q+1}\ldots{\sigma_{m}},\sigma_{m-q-p+1}\ldots\sigma_{m-q})$
\end{itemize}
	for all strings $\sstring{m}\in\Sigma^{+}$, is injective.
\end{definition}
	Developing applications based on $(p,q)$-adaptive codes is not a subject of this paper. The concept is presented here
	just to show how much flexibility we get when using various generalizations of adaptive codes. Let us give an example.
\begin{example}
	Let $\Sigma=\{\ttup{a},\ttup{b}\}$, $\Delta=\{0,1\}$ be two alphabets, and
	$c:\Sigma^{2}\times\Sigma^{\leq{1}}\rightarrow\Delta^{+}$ a function given as in the table below.
	One can verify that $\overline{c}$ is injective, and according to Definition \ref{def:pqac}, $c$ is
	an $(1,2)$-adaptive code.
\begin{table}[htbp]
\caption{An $(1,2)$-adaptive code.}
\begin{center}
\begin{tabular}{|c|c|c|c|c|}
\hline
$\Sigma^{2}\backslash\Sigma^{\leq{1}}$	& $\ttup{a}$ & $\ttup{b}$	& $\lambda$	\\ \hline
$\ttup{aa}$ 					    	& 0          & 11 	        & 00	   	\\ \hline
$\ttup{ab}$					        	& 10         & 101	        & 11       	\\ \hline
$\ttup{ba}$					        	& 111        & 01	        & 10	   	\\ \hline
$\ttup{ba}$					        	& 110        & 00	        & 01	   	\\ \hline
\end{tabular}
\end{center}
\end{table}
\hspace{0pt}
\newline
	Let $x=\ttup{ababa}\in\Sigma^{+}$ be an input data string. 
	Using the definition above, we encode $x$ by
\begin{center}
	$\overline{c}(x)=c(\ttup{ab},\lambda)c(\ttup{ba},\ttup{a})c(\ttup{ab},\ttup{b})c(\ttup{ba},\ttup{a})=11111101111$.
\end{center}
\end{example}
\section{Adaptive codes and time-varying codes}
	Time-varying codes have been recently introduced in \cite{tme:1} as a proper extension of L-codes \cite{m:s:w}.
	Intuitively, a time-varying code associates a codeword to the symbol being encoded depending on its position in the input data string.
	The connection to gsm-codes and SE-codes has been also discussed in \cite{tme:1}.
	Several characterizations results for time-varying codes can be found in \cite{tmte:1}. Let us now give a formal definition.
\begin{definition}
\label{def:tvc}
	Let $\Sigma$ and $\Delta$ be two alphabets. A function
	$c:\Sigma\times\mathbb{N}^{*}\rightarrow\Delta^{+}$ is called \emph{time-varying code} if its unique
	homomorphic extension $\overline{c}:\Sigma^{*}\rightarrow\Delta^{*}$, given by:
\begin{itemize}
\item $\overline{c}(\lambda)=\lambda$,
\item $\overline{c}(\sstring{m})=c(\sigma_{1},1)c(\sigma_{2},2)\ldots{c(\sigma_{m},m)}$
\end{itemize}
	for all strings $\sstring{m}\in\Sigma^{+}$, is injective.
\end{definition}
	{\bf Motivation.} This section is intended to introduce a new class of variable-length codes, called
	{\it adaptive time-varying codes}. Combining adaptive codes with time-varying codes can be useful when the input string
	consists of substrings with different characteristics.
	Let $x=u_{1}u_{2}\ldots{u_{t}}\in\Sigma^{+}$ be an input string, where $u_{1},u_{2},\ldots,u_{t}$ are substrings with
	different characteristics. Instead of associating an adaptive code to $x$, it is desirable to associate an adaptive code
	to each substring $u_{i}$.
	For sure, this technique can be exploited further in data compression to improve the results.
	Combining adaptive codes with time-varying codes leads to the following encoding mechanism: the codeword associated to the
	current symbol being encoded depends not only on the previous symbols in the input string, but also on the position of the current
	symbol in the input string. A formal definition is given below.
\begin{definition}
\label{def:tvac}
	Let $\Sigma$ and $\Delta$ be alphabets. A function
	$c:\Sigma\times\Sigma^{\leq{n}}\times\mathbb{N}^{*}\rightarrow\Delta^{+}$ is called \emph{adaptive time-varying code of order $n$}
	if its unique homomorphic extension $\eac$, given by:
\begin{itemize}
\item $\overline{c}(\lambda)=\lambda$,
\item $\overline{c}(\sstring{m})=$
	$c(\sigma_{1},\lambda,1)$
	$c(\sigma_{2},\sigma_{1},2)$
	$\ldots$
	$c(\sigma_{n-1},\sstring{n-2},n-1)$
	\newline
	$c(\sigma_{n},\sstring{n-1},n)$
	$c(\sigma_{n+1},\sstring{n},n+1)$
	$c(\sigma_{n+2},\sigma_{2}\sigma_{3}\ldots\sigma_{n+1},n+2)$
	\newline
	$c(\sigma_{n+3},\sigma_{3}\sigma_{4}\ldots\sigma_{n+2},n+3)\ldots$
	$c(\sigma_{m},\sigma_{m-n}\sigma_{m-n+1}\ldots\sigma_{m-1},m)$
\end{itemize}
	for all strings $\sstring{m}\in\Sigma^{+}$, is injective.
\end{definition}
\begin{example}
	Let $\Sigma=\{\ttup{a},\ttup{b}\}$, $\Delta=\{0,1\}$ be two alphabets, and let
	$c:\Sigma\times\Sigma^{\leq{2}}\times\mathbb{N}^{*}\rightarrow\Delta^{+}$ be a function given by:
\begin{displaymath}
	c(\sigma,u,i)=
	\left\{
		\begin{array}{ll}
			{\it zero[i]}				& \textrm{if $\sigma=\ttup{a}$.}\\
			{\it one[i]}				& \textrm{if $\sigma=\ttup{b}$.}
		\end{array}
	\right.
\end{displaymath}
	for all $(\sigma,u,i)\in\Sigma\times\Sigma^{\leq{2}}\times\mathbb{N}^{*}$, where
\begin{itemize}
\item ${\it zero[i]}=\underbrace{00\ldots{0}}_{i}$,
\item and ${\it one[i]}=\underbrace{11\ldots{1}}_{i}$.
\end{itemize}
	One can verify that $\overline{c}$ is injective, and according to Definition \ref{def:tvac}, $c$ is an adaptive
	time-varying code of order two. For example, the string $x=\ttup{abaa}$ is encoded by
\begin{displaymath}
	\overline{c}(x)=c(\ttup{a},\lambda,1)c(\ttup{b},\ttup{a},2)c(\ttup{a},\ttup{ab},3)c(\ttup{a},\ttup{ba},4)=0110000000.
\end{displaymath} 
\end{example}
\section{Conclusions and further work}
	Adaptive codes associate variable-length codewords
	to symbols being encoded depending on the previous symbols in the input data string.
	This class of codes has been presented in \cite{t:2} as a new class of non-standard variable-length codes.
	Generalized adaptive codes (GA codes, for short) have been also presented in \cite{t:2}, not only as a
	new class of non-standard variable-length codes, but also as a natural generalization of adaptive codes
	of any order.

	In this paper, we contributed the following results.
	First, we proved that adaptive Huffman encodings and Lempel-Ziv encodings
	are particular cases of encodings by GA codes (sections 3 and 4).
	In section 5, we proved that any $(n,1,m)$ convolutional code satisfying a certain condition can be modelled as an adaptive code
	of order $m$.
	This result was exploited further in section 6, where an efficient cryptographic scheme based on convolutional codes is described.
	An insightful analysis of this cryptographic scheme was provided in the same section.
	In sections 7 and 8, we extended adaptive codes to $(p,q)$-adaptive codes, and presented a new class of variable-length codes, called
	adaptive time-varying codes.

	Further work in this area is intended to establish new interesting connections between adaptive codes and other classes of codes,
	along with showing their effectiveness in concrete applications.
	Future directions related to adaptive codes also include the data compression algorithms recently presented in \cite{t:2}.
	For example, combining the extensions described in sections 7 and 8 with the algorithms presented in \cite{t:2} may lead
	to better results.
%\begin{acknowledgements}
%	The author would like to thank the anonymous referee for pointing out the connection between adaptive codes
%	and convolutional encoding.
%\end{acknowledgements}

\bibliographystyle{fundam}
\begin{small}

\end{small}

%%%%%%%%%%%%%%%%%%%%%%%%%%%%%%%%%%%%%%%%%%%%%%%%%%%%%%%%%%%%%%%%%%%%%%

\end{document}